\documentclass[twocolumn,prb]{revtex4}
\usepackage{amsfonts}
\usepackage[T1]{fontenc}
\usepackage{amsmath,amsbsy,amssymb,graphicx}
\usepackage{times}
\usepackage{mathrsfs}

\begin{document}

\title{Magnetic-Field Induced Semimetal in Topological Crystalline Insulator
Thin Films}
\author{Motohiko Ezawa}

\begin{abstract}
We investigate electromagnetic properties of a topological crystalline
insulator (TCI) thin film under external electromagnetic fields. The TCI
thin film is a topological insulator indexed by the mirror-Chern number. It
is demonstrated that the gap closes together with the emergence of a pair of
gapless cones carrying opposite chirarities by applying in-plane magnetic
field. A pair of gapless points have opposite vortex numbers. This is a
reminiscence of a pair of Weyl cones in 3D Weyl semimetal. We thus present
an a magnetic-field induced semimetal-semiconductor transition in 2D
material. This is a giant-magnetoresistance, where resistivity is controlled
by magnetic field. Perpendicular electric field is found to shift the
gapless points and also renormalize the Fermi velocity in the direction of
the in-plane magnetic field.
\end{abstract}

\maketitle

\address{{\normalsize Department of Applied Physics, University of Tokyo, Hongo
7-3-1, 113-8656, Japan }}



\textit{Introduction:} Topologically stable states such as topological
insulator are among the most exciting topics in modern condensed matter
physics. Topological crystalline insulator (TCI) is a topological insulator
protected by the crystal symmetry\cite%
{Fu,Hsieh,LiuFu,FangBer,LiuFu04,Okada,TCIOpt}. Its experimental realizations
in Pb$_{x}$Sn$_{1-x}$Te excite studies of TCI\cite{Ando,Xu,Dz}. A thin film
made of TCI provides us with a new platform of 2D electron system\cite%
{LiuFu,EzawaTCIfilm,Yamakage}. When the film is thin enough, the gap opens
due to hybridization between the front and back surfaces, and it turns the
system into a topological insulator. The TCI thin film is a topological
insulator indexed by the mirror-Chern number\cite{TeoMirror,Takahashi}. Very
recently, a TCI thin film made of SnTe was experimentally manufactured\cite%
{Taskin}.

Weyl semimetal\cite{Murakami,
HosurRev,Weh,Vafek,Burkov,Hosur,Hal,Wan,Zyu,Sekine} has recently been found
to be topologically robust in three dimensions (3D). The emergence of a Weyl
semimetal is always accompanied by a pair of Weyl cones with opposite
chiralities subject to the fermion doubling theorem\cite{Nielsen}. Each Weyl
cone has a gapless point in momentum space, carrying the opposite monopole
number. A pair of monopoles cannot annihilate each other dynamically, since
they are parts of the ground-state texture of the Berry curvature. The
semimetal is topologically stable provided two Weyl cones are separated.
Indeed, when two Weyl cones meet head-on by controlling system parameters,
they disappear and the gap opens in the system. These are the basic features
of the 3D Weyl semimetal.

We investigate the band structure and the topological property of TCI thin
film by applying the in-plane magnetic field and the perpendicular electric field. 
As these external fields are increased, the gap reduces and closes,
forming a gapless Dirac cone at certain critical fields. Then the gapless
Dirac cone splits into a pair of gapless cones with opposite chiralities.
They are akin to a pair of Weyl cones in 3D. Indeed, they carry the opposite
vortex numbers. We may call the gapless cone (point) "Weyl" cone (point) in
2D based on the similarity. We find a flat band emerges to connect them in a
nanoribbon, as is a reminiscence of the Fermi arc\cite{Wan} connecting the
two Weyl points in the surface of a Weyl semimetal. It is to be emphasized
that the emergence of the Weyl points is solely due to the in-plane magnetic
field, while the electric field only shifts the position of the Weyl points
and renormalizes the Fermi velocity. When we change the direction of
in-plane magnetic field, the gap remains open and the positions of the Weyl
points rotate in parallel to the magnetic field direction. The pair of Weyl
points never annihilate each other provided they are separated by the
in-plane magnetic field. Thus we have presented a magnetic-field induced
semimetal-semiconductor transition in 2D material.

\textit{Topological Crystalline Insulator Thin Film: }Gapless Dirac cones
emerge on the surface of a topological insulator. We consider the [0,0,1]
TCI surface, where there are gapless Dirac cones at the $X$ and $Y$ points.
When the thickness is very thin, the gap opens due to hybridization between
the front and back surfaces. We explicitly investigate the low-energy
physics near the Fermi energy around the $X$ point, but the same analysis
can be carried out also around the $Y$ point. It is well known that the
low-energy physics in the vicinity of the Dirac point is described precisely
by the Dirac theory. Hence we are able to present analytic formulas.
Nevertheless, we also carry out numerical studies based on the tight-binding
model to confirm analytical results.

The effective low-energy Hamiltonian of the TCI thin film has been derived
in the vicinity of the $X$ point\cite{LiuFu}. It is expressed in terms of $%
4\times 4$ matrices,%
\begin{equation}
H_{0}=\left[ v_{x}k_{x}\sigma _{x}-v_{y}k_{y}\sigma _{y}\right] \tau
_{y}+m\tau _{x},  \label{DiracHamil}
\end{equation}%
where $\sigma =(\sigma _{x},\sigma _{y},\sigma _{z})$ and $\tau =(\tau
_{x},\tau _{y},\tau _{z})$ represent the spin and surface degrees of
freedom; $v_{i}$ and $k_{i}$ are the Fermi velocity and the momentum into
the $i$-direction; $m\tau _{x}$ represents the tunnelling term between the
two surfaces. We have set $\hbar =1$ for simplicity.\ The Hamiltonian $H_{0}$
has the mirror symmetry, $MH_{0}(\mathbf{k})M^{-1}=H_{0}(\mathbf{k})$, with
the generator $M=-\frac{i}{2}\sigma _{z}\tau _{x}$.

\begin{figure}[t]
\centerline{\includegraphics[width=0.45\textwidth]{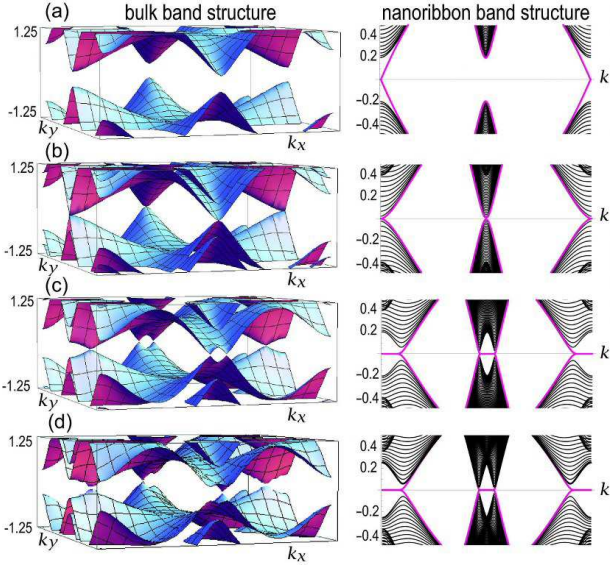}}
\caption{Band structure of bulk and nanoribbon of the TCI thin film based on
the tight-binding model. (a) Without external field ($B_{x}=0,B_{y}=0$). (b)
At the phase transition point ($B_{x}=m,B_{y}=0$), where the band gap
closes. (c) In the semimetallic phase ($B_{x}=2m,B_{y}=0$). (d) In the
semimetallic phase ($B_{x}=2m,B_{y}=m$). We have set $m=0.2$.}
\label{FigRibbon}
\end{figure}

Without the external fields, the Hilbert space is divided by the eigenvalues
($M=\pm i$) of the mirror operator\cite{TeoMirror,Takahashi}. The
mirror-Chern number $C_{M}$\ is defined by the difference of the Chern
numbers\ in these two sectors. The total Chern number is zero ($C=0$) and
the mirror-Chern number is given by $C_{M}=\frac{1}{2}\text{sign}(m)$ for
each cone. It is a mirror-Chern insulator. When we analyze a nanoribbon
based on the tight-binding model\cite{EzawaTCIfilm}, gapless edge modes
appear, as signals the topological nature of the bulk [Fig.\ref{FigRibbon}%
(a)].

We introduce the external field terms%
\begin{equation}
H_{\text{ext}}=E_{z}\tau _{z}+B_{x}\sigma _{x}+B_{y}\sigma _{y}.
\end{equation}%
The first term is induced by applying electric field perpendicular to the
TCI thin film. The in-plane Zeeman terms are induced by applying in-plane
magnetic field.

We explore the system with $H=H_{0}+H_{\text{ext}}$. The band structure
changes as a function of the external fields. The band gap is located at $%
k_{x}=k_{y}=0$, where the energy spectrum reads 
\begin{equation}
E=\pm \left\vert \sqrt{B_{x}^{2}+B_{y}^{2}}\pm \sqrt{m^{2}+E_{z}^{2}}%
\right\vert .
\end{equation}%
The gap closes when $B_{x}^{2}+B_{y}^{2}=m^{2}+E_{z}^{2}$, where a
topological phase transition occurs [Fig.\ref{FigRibbon}(b)].

Before the gap closes the system is an insulator. However, it is no longer a
mirror-Chern insulator since the mirror symmetry is broken. When we examine
a nanoribbon based on the tight-binding model\cite{EzawaTCIfilm}, edge modes
are gapped. The mirror-Chern number $C_{M}$ becomes a continuous function of 
$E_{z}$, $B_{x}$ and $B_{y}$ and is no longer quantized. For instance, we
find\cite{EzawaSTP} 
\begin{equation}
\mathcal{C}_{M}=\frac{m}{2\sqrt{m^{2}+E_{z}^{2}}},
\end{equation}%
when we apply only $E_{z}$. It is reduced to the quantized value, $C_{M}=%
\frac{1}{2}$sign$(m)$, in the limit $E_{z}=0$.

\begin{figure}[t]
\centerline{\includegraphics[width=0.45\textwidth]{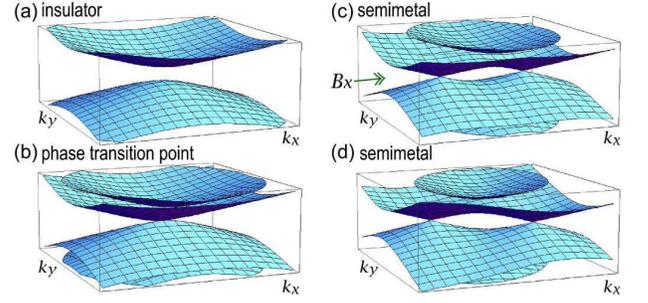}}
\caption{Band structure near the $X$ point based on the 4-band theory. (a)
Without external field ($B_{x}=0,B_y=0$). (b) At the phase transition point (%
$B_{x}=m,B_y=0$), where the band gap closes. (c) In the semimetallic phase ($%
B_{x}=2m,B_y=0$). (d) In the semimetallic phase ( $B_{x}=2m,B_y=m$). We have
set $m=0.2$.}
\label{FigWeyl}
\end{figure}

\textit{Magnetic field induced Semimetal: }We investigate the TCI thin film
after the phase transition point, where the gap closes [Fig.\ref{FigRibbon}%
(c)]. We may choose the direction of the in-plane magnetic field as the $x$%
-axis without loss of generality. To study the phenomenon analytically we
examine the energy spectrum of the Dirac theory, 
\begin{equation}
E=\pm \sqrt{v_{y}^{2}k_{y}^{2}+\left( \sqrt{%
v_{x}^{2}k_{x}^{2}+m^{2}+E_{z}^{2}}\pm B_{x}\right) ^{2}}.  \label{Energ4}
\end{equation}%
We show the band structure in Fig.\ref{FigWeyl}. The phase transition point
is $B_{x}^{2}=m^{2}+E_{z}^{2}$. Beyond the point, a gapless Dirac cone is
decomposed into two Weyl cones located at $(k_{x},k_{y})=(\pm k_{x}^{X},0)$
as in Fig.\ref{FigWeyl}(c) with $k_{x}^{X}=v_{x}^{-1}\sqrt{%
B_{x}^{2}-m^{2}-E_{z}^{2}}$. We also refer to these Weyl points as the $%
X_{\pm }$ points. It is to be emphasized that the decomposition is made
possible by the in-plane magnetic field. We also find that flat edge modes
appear connecting the two Weyl points in a nanoribbon [Fig.\ref{FigRibbon}%
(c)], which would correspond to the Fermi arc\cite{Wan} connecting the two
Weyl points in the surface of a 3D Weyl semimetal.

We employ the effective $2\times 2$ Hamiltonian by taking only two bands
nearest to the Fermi energy. We may derive it in the second order
perturbation theory around the $X_{\pm }$ point as follows. Since the
quantization axis is $\sigma _{x}$\ and $\tau _{x}$\ at the $\Gamma $\
point, it is convenient to make the cyclic rotation of the Pauli matrices
and diagonalize them. In the new basis the Hamiltonian reads 
\begin{equation}
H_{0}=\left[ v_{x}k_{x}\sigma _{z}-v_{y}k_{y}\sigma _{x}\right] \tau
_{x}+m\tau _{z}+E_{z}\tau _{x}+B_{x}\sigma _{z},
\end{equation}%
which is explicitly written as 
\begin{equation}
H=\left( 
\begin{array}{cc}
H_{1} & T \\ 
T^{\dagger } & H_{2}%
\end{array}%
\right) ,  \label{Hamil4}
\end{equation}%
with 
\begin{align}
H_{1}& =(m-B_{x})\sigma _{z}-v_{y}k_{y}\sigma _{x}, \\
H_{2}& =(-m-B_{x})\sigma _{z}-v_{y}k_{y}\sigma _{x}, \\
T& =-v_{x}k_{y}\sigma _{z}-iE_{z}\sigma _{y}.
\end{align}%
The dominant term is $H_{1}$. In the second-order perturbation theory, we
obtain the effective Hamiltonian 
\begin{align}
H_{\text{eff}}& =H_{1}-T^{\dagger }H_{2}^{-1}T  \notag \\
& =(B_{x}-m-\frac{(m+B_{x})(v_{x}^{2}k_{x}^{2}+E_{z}^{2})}{%
(m+B_{x})^{2}+v_{y}k_{y}^{2}})\sigma _{z}  \notag \\
& -v_{y}k_{y}(1+\frac{v_{x}^{2}k_{x}^{2}+E_{z}^{2}}{%
(m+B_{x})^{2}+v_{y}k_{y}^{2}})\sigma _{x}.
\end{align}%
By neglecting higher order terms in $k_{y}$\ and changing the Pauli matrices
inversely, we obtain the effective Hamiltonian 
\begin{equation}
H_{\text{eff}}=\left( B_{x}-m-\frac{v_{x}^{2}k_{x}^{2}+E_{z}^{2}}{m+B_{x}}%
\right) \sigma _{x}-v_{y}k_{y}\sigma _{y},  \label{Heff}
\end{equation}%
with the energy spectrum being 
\begin{equation}
E=\pm \sqrt{\left( \frac{v_{x}^{2}k_{x}^{2}+m^{2}+E_{z}^{2}-B_{x}^{2}}{%
m+B_{x}}\right) ^{2}+v_{y}^{2}k_{y}^{2}}.
\end{equation}%
It agrees with (\ref{Energ4}) up to the order of $k_{x}^{4}$.

At the transition point $B_{x}^{2}=m^{2}+E_{z}^{2}$ [Fig.\ref{FigWeyl}(b)],
the energy spectrum is highly anisotropic. The dispersion is Schr\"{o}%
dinger-like in the $k_{x}$ direction, and Dirac-like in the $k_{y}$
direction.

\begin{figure}[t]
\centerline{\includegraphics[width=0.25\textwidth]{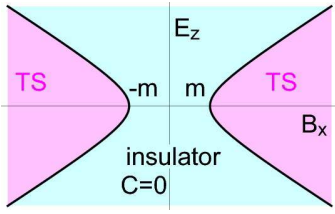}}
\caption{Topological phase diagram. In the $B_{x}$-$E_{z}$ plane, with the
topological semimetal (TS) and the insulator ($C=0$). }
\label{FigBxzPhase}
\end{figure}

Beyond the transition point $B_{x}^{2}>m^{2}+E_{z}^{2}$ [Fig.\ref{FigWeyl}%
(c)], we rewrite the Hamiltonian (\ref{Heff}) as%
\begin{equation}
H_{\text{eff}}=-\frac{v_{x}^{2}\left( k_{x}-k_{x}^{X}\right) \left(
k_{x}+k_{x}^{X}\right) }{m+B_{x}}\sigma _{x}-v_{y}k_{y}\sigma _{y}.
\label{Effec22}
\end{equation}%
We may approximate $k_{x}\pm k_{x}^{X}=\pm 2k_{x}^{X}$ around the $X_{\pm }$
point, 
\begin{equation}
H_{\text{eff}}^{X_{\pm }}=\mp \tilde{v}_{x}\left( k_{x}\mp k_{x}^{X}\right)
\sigma _{x}-v_{y}k_{y}\sigma _{y},  \label{EffecX}
\end{equation}%
with the renormalized velocity $\tilde{v}_{x}=2v_{x}^{2}k_{x}^{X}/(m+B_{x})$%
. The energy spectrum in the vicinity of the $X_{\pm }$ points is%
\begin{equation}
E_{\pm }=\pm \sqrt{\tilde{v}_{x}^{2}\tilde{k}_{x}^{2}+v_{y}^{2}k_{y}^{2}}.
\end{equation}%
The role of $E_{z}$ is to shift the position of the Weyl points and
renormalize the Fermi velocity.

It is easy to consider the general in-plane field $B_{x}\neq 0$\ and $%
B_{y}\neq 0$\ when $E_{z}=0$. We find that the gap closes at $%
(k_{x}^{X},k_{y}^{X})$\ and $(-k_{x}^{X},-k_{y}^{X})$, where%
\begin{equation}
(k_{x}^{X},k_{y}^{X})=(v_{x}^{-1}\left\vert B_{x}\right\vert
,v_{y}^{-1}\left\vert B_{y}\right\vert )/\sqrt{1-m^{2}/B_{\Vert }^{2}},
\end{equation}%
with $B_{\Vert }^{2}=B_{x}^{2}+B_{y}^{2}$. The positions of Weyl cones are
parallel to the direction of the magnetic field. We have also confirmed the
results by calculating the band structure of the bulk and a nanoribbon based
on the tight-binding model. Namely the gap remains closed for an arbitral
direction of the in-plane field, as we show an example in Fig.\ref{FigRibbon}%
(e).

\begin{figure}[t]
\centerline{\includegraphics[width=0.45\textwidth]{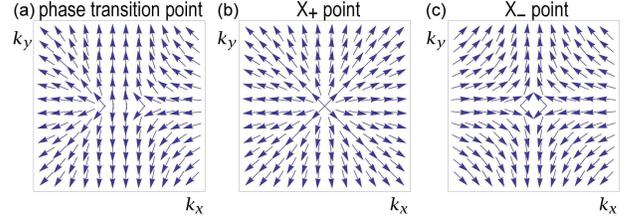}}
\caption{Spin direction based on the 2-band theory. (a) In the topological
insulator ($M_{x}=0$). (b) At the transition point ($M_{x}=m$). (c) In the
topological semimetal ($M_{x}=2m$), where a hedgehog structure is found at
the $X_{+}$ point, and an anti-hedgehog structure at the $X_{-}$ point. The
vortex number is nontrivial at these points. }
\label{FigDiracSpin}
\end{figure}

We discuss the topological stability of the $X_{\pm }$\ points. The
Hamiltonian is of the form 
\begin{equation}
H_{\text{eff}}=R(n_{x}\sigma _{x}+n_{y}\sigma _{y}),
\end{equation}%
where $n_{x}$ and $n_{y}$ are normalized fields subject to $%
n_{x}^{2}+n_{y}^{2}=1$. It has two eigen-spinors with the eigen-energy $%
E_{\pm }=\pm |R|$. The state corresponding to the filled band is given by
the spinor 
\begin{equation}
|S\rangle =\frac{1}{\sqrt{2}}\left( 
\begin{array}{c}
e^{-i\phi /2} \\ 
-e^{+i\phi /2}%
\end{array}%
\right) ,
\end{equation}%
when we parametrize $(n_{x},n_{y})=(\cos \phi ,\sin \phi )$. The spin of the
state is given by $s_{i}=\langle S|\sigma _{i}|S\rangle $, which we show in
Fig.\ref{FigDiracSpin}. We note that $s_{i}=-n_{i}$. The spin direction
forms a hedgehog structure in the vicinity of the $X_{+}$ point, while it
forms an anti-hedgehog structure in the vicinity of the $X_{-}$ point. There
are a source and a sink of the spin flow at these points. They are described
by a topological charge.

We may define the vortex number for the spin texture by\cite{EzawaMeron} 
\begin{equation}
Q=\frac{1}{2\pi }\oint dk_{\alpha }\,[s_{x}(\mathbf{k})\partial _{k_{\alpha
}}s_{y}(\mathbf{k})-s_{y}(\mathbf{k})\partial _{k_{\alpha }}s_{x}(\mathbf{k}%
)],  \label{NumbeVorte}
\end{equation}%
where the integration is carried out along the boundary of the Brillouin
zone. It is trivial to see%
\begin{equation}
Q=\frac{1}{2\pi }\oint dk_{\mu }\,\partial _{k_{\mu }}\phi .
\label{VortexNumbe}
\end{equation}%
In general, it yields $Q=0,\pm 1,\pm 2,\cdots $, because $\phi $ is defined
only modulo $2\pi $. For the specific field configuration given in (\ref%
{EffecX}), we find $Q=\pm 1$ for the Weyl cones at $X_{\pm }$. These values
are topologically stable because any perturbation cannot change the
quantized value of the topological charge $Q$. We may also evaluate the
topological number for the spin configuration at the phase transition point
[Fig.\ref{FigDiracSpin}(a)] and also before the phase transition point (i.e.
in the insulator phase) to find that $Q=0$. We may interpret that a pair
creation of Weyl cones with $Q=\pm 1$ occurs from the topological trivial
Dirac cone with $Q=0$ at the phase transition point. It is interesting to
note that a flat edge mode appear to connect the two $X_{\pm }$ points in a
nanoribbon [Fig.\ref{FigRibbon}(c)]. Note that the total topological charge (%
\ref{NumbeVorte}) is zero ($Q=0$) both before and after the phase
transition. Nevertheless, the semimetallic phase is stable topologically due
to the presence of a pair of two Weyl cones with $Q=\pm 1$\ generated by the
in-plane field, precisely as in the 3D Weyl semimetal.

The same analysis is carried out for the Dirac cone at the $Y$ point, from
which a pair of Weyl cones emerge located at the $Y_{\pm }$ point under
in-plane magnetic field.

\textit{Discussions: }We have demonstrated that a semimetallic phase emerges
in a TCI thin film by applying in-plane magnetic field. The semimetallic
phase is characterized by the existence of a pair of gapless Weyl cones
as in the case of the 3D Weyl semimetal. 
In conclusion, we have proposed a magnetic-field induced
semimetal-semiconductor transition in 2D material. This is a
giant-magnetoresistance, where resistivity is controlled by magnetic field.
The transition between the insulator phase and the semimetallic phase will
be experimentally detectable by electric transport measurement. The TCI thin
film has already been manufactured\cite{Taskin}. Our finding will open a way to
magneto-nanoelectronics based on the TCI thin film.

I am very much grateful to N. Nagaosa, Y. Ando and Y. Tanaka for helpful
discussions on the subject. This work was supported in part by Grants-in-Aid
for Scientific Research from the Ministry of Education, Science, Sports and
Culture No. 25400317.

\end{document}